# Search for supersolid $^4$He in neutron scattering experiments at ISIS

O. Kirichek

*ISIS Facility, STFC, Rutherford Appleton Laboratory, Harwell, Didcot, OX11 0QX, UK*

E-mail: o.kirichek@rl.ac.uk

**Abstract**. The observation of a non-classical rotation inertia (NCRI) fraction in bulk solid $^4$He by M. Chan and E. Kim [1, 2] attracted significant interest as a possible manifestation of supersolid state of matter. Despite numerous experimental and theoretical studies inspired by this observation, an explicit explanation for this phenomenon is still missing. Neutron scattering experiments on solid helium may help to shed light on the physical grounds of NCRI and answer the question on whether this phenomenon could be caused by Bose-Einstein Condensation. In this paper we are going to discuss the results obtained in experiments involving neutron scattering on solid $^4$He. Microscopic quantitative data such as mean kinetic energy, mean square momentum and mean square displacement of helium atoms as well as the lattice parameter have been obtained for the first time for solid $^4$He in temperature range 70mK – 500 mK. No change was seen in the single atom kinetic energy (within statistical error better than 1%) as well as change in the lattice parameter (within 0.03%). The mean square displacement did not change in the region of expected supersolid transition either. All these results suggest that the NCRI transition is quite different from the superfluid transition in liquid $^4$He.

## 1. Introduction

Following the remarkable observation [1, 2] of a non-classical rotational inertia (NCRI) in bulk solid $^4$He, which may be a possible manifestation of "supersolidity", several independent groups which also used torsional oscillator experimental technique have confirmed the result [3]. However, a number of experiments based on other methods [3] have shown no anomaly in the possible supersolid transition area. A lack of clear theoretical explanation, in combination with the controversial experimental results make the situation with NCRI quite ambiguous. Neutron scattering experiments on solid helium may help to shed light on the physical grounds of NCRI.

In early neutron diffraction measurements of the $^4$He solidified in a mesoporous glass [4] no change of the diffraction pattern that can be related to the onset of supersolid transition has been observed. The sample was studied along the isochoric path, where about 30% of the pore filling solidify in the *bcc* structure at 1.7K and remain in this state down to 60 mK. The other part of the pore filling was either liquid or an amorphous solid. These results raised concerns about a presence of liquid or amorphous helium fraction in the porous media, which can significantly change the mechanical properties of the sample.

Recently the excitation spectrum of bulk hcp solid $^4$He has been directly probed by inelastic neutron scattering [5]. In the experiment a branch of delocalized excitations, with both longitudinal and transverse dispersion that coexist with acoustic phonons, has been observed. While at high accuracy there was no change in the density of states between 0.15 meV and 5.5 meV upon cooling through the supersolid transition, the new collective modes remain an intriguing aspect of solid helium.

In this paper we are going to discuss the results obtained in neutron scattering on solid $^4$He experiments. The experiments have been carried out on three instruments: VESUVIO, MARI and MAPS at the ISIS neutron source. All samples, consisting of either single crystal or polycrystalline have been grown using the "blocked capillary technique". Microscopic quantitative data such as the mean kinetic energy, mean square momentum, mean square displacement of helium atoms as well as the lattice parameter have been obtained for the first time for solid $^4$He in temperature range 70mK – 500 mK.

## 2. Measurement of the kinetic energy and lattice constant on the VESUVIO spectrometer

The VESUVIO spectrometer uses eV neutrons to measure atomic kinetic energies and momentum distributions. The instrument detects only scattered neutrons with the energies of 4.908 ± ~0.15 eV. The time of flight measurement is then used to determine the incident neutron energy and hence the momentum and energy transfer in the scattering process. Measurements of atomic kinetic energies by neutron scattering are possible, as at these high incident neutron energies the impulse approximation (IA) [6] is valid.

Three different samples were prepared to test the effects of crystal quality and $^3$He impurity concentration, which are thought to be important for the observation of supersolidity [3]. The sample growing procedure is presented in Fig.1 by dotted line *a*.

In order to prepare sample A, helium was initially condensed into the cell at the temperature ~2K, after which the cell temperature was increased and controlled at 2.9K. The $^4$He pressure was then raised slowly up to 70 bars. At that point the temperature was quickly dropped by 100 mK and the capillary was blocked by a solid helium plug. It was then decreased by 100 mK each hour down to 2K where it was kept for another 8 Hours. The sample was then cooled to the base temperature of the dilution refrigerator. At temperatures less than 1K the sample pressure was 40.6 bar. The scarcity of observable Bragg peaks and the absence of (002) peak was consistent with the presence of an hcp single crystal with *c* axis along the axis of the cylinder.

For the preparation of sample B the sample was warmed to the liquid phase at 3.4K and 70 bars and then cooled quickly to the base temperature. To prepare sample C, the helium was removed from the cell and replaced with a $^3$He-$^4$He mixture containing 10ppm of $^3$He. The sample was then cooled rapidly from 3.5K at 70 bars, down to the base temperature. Diffraction measurements showed many Bragg peaks suggesting that both samples B and C were polycrystalline and had the same density as sample A.

As it follows from neutron Compton Scattering data obtained in the experiment [7], the single atom kinetic energy $\kappa$ of solid hcp $^4$He at temperatures between 0.07 and 0.4K and a pressure of 40 bars demonstrated no change within the statistical error of ~2% for all samples. The lattice constant was also measured and again was found to be constant to within 0.05%.

## 3. Search for Bose-Einstein Condensation in Solid $^4$He

In liquid helium, Bose-Einstein condensation (BEC) and superfluidity are observed together. Indeed superflow can be shown to follow from BEC [8]. In the liquid, BEC is observed as a macroscopic occupation of the $k_0$ state in $n_\mathbf{k}$, as expected for a translationally invariant system. Therefore for some models of superflow in solid helium, such as gas of vacancies, the BEC should appear as a macroscopic occupation of $k_0$ in $n_k$. In this context we look for an enhancement of $n_k$ at $k_0$ below $T_c$. We also look for a change in the shape of $n_k$ below $T_c$.

The solid $^4$He sample studied in this experiment was grown following the same procedure as in VESUVIO experiment sample B (see Fig. 1 dotted line *a*). Commercial grade purity $^4$He (0.3 ppm $^3$He) was introduced into a cylindrical Al sample cell at a temperature of 3 K to a pressure of 70 bars. At this pressure, the temperature was reduced until a solid formed in the capillary and blocked the cell. The block was observed at pressure 69.8 bars and temperature 2.79 K. The blocked cell was further cooled and neutron inelastic scattering measurements at high momentum transfer were collected in the solid hcp phase at 80, 120, 300, and 500 mK on the MARI spectrometer at the ISIS neutron facility.

To obtain a condensate fraction from experimental neutron scattering data, we assumed a model *n(k)* of the form:

$$n(k) = n_0 \delta(k) + (1 - n_0) \cdot n^*(k)$$

where *n\*(k)* is the momentum distribution of the atoms above the condensate in the $k > 0$ states.

As the result of the analysis [9] we have determined that the BEC condensate fraction $n_0 = (-0.10 \pm 1.20)\%$ at T = 80 mK below $T_c$ = 200 mK is consistent with zero. T = 80 mK is somewhat above, but close to the temperature T = 50 mK at which $\rho_s$ reaches its maximum value [1, 2]. We also found no change in the shape of the atomic momentum distribution on the crossing $T_c$.

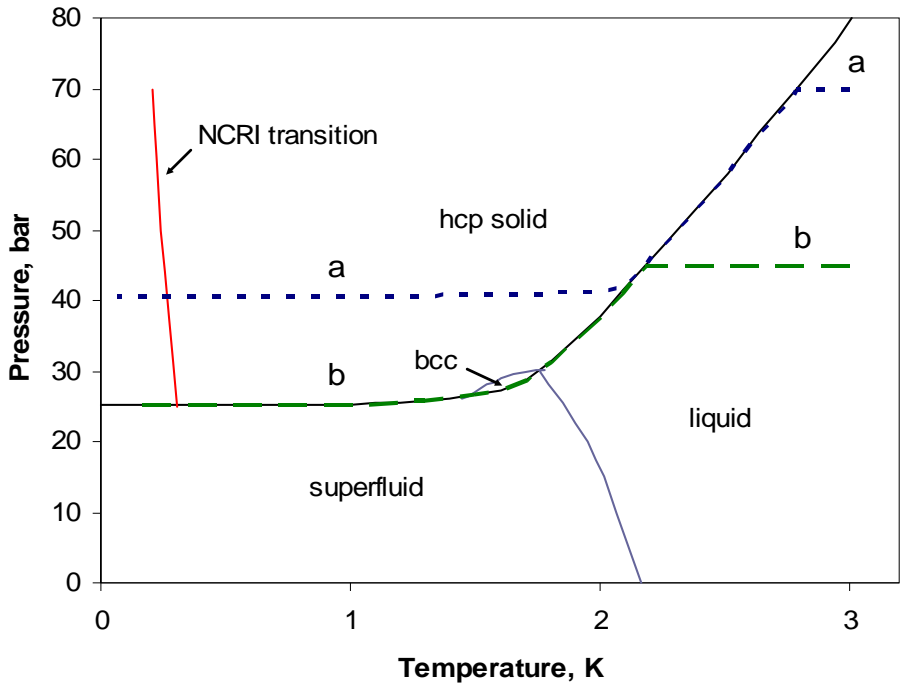

Fig.1 The sample growing procedure presented at $^4$He phase diagram

## 4. The mean-square displacement of solid $^4$He crystal down to 140 mK.

In this experiment the sample chamber supported the growth of multiple crystallites with different orientations, as opposed to a single crystal. The sample growing procedure is presented in Fig.1 by dash line *b*. The sample examined here was formed by starting at pressure 45 bars and temperature 2 K. Solid $^4$He was then grown using the "blocked capillary technique," in which a solid plug was formed in the capillary as the system cooled so that the molar volume of the sample remained constant as it cooled along the melting curve. This means that our sample passed through the region of bcc phase on the melting curve which presumably contributed to the formation of the strained crystallites.

The neutron data were collected using the MAPS time-of-flight spectrometer at the ISIS neutron facility. The mean-square displacement $\langle u^2 \rangle$ and the lattice parameter have been determined for the same crystallite from the diffraction patterns obtained over the temperature range 140–800 mK [10].

The principal result of this experiment was that the mean-square displacement $\langle u^2 \rangle$ does not change over the temperature range 140–800 mK. Within the precision of these data, there is no apparent temperature dependence at all, i.e., there is no indication that the supersolid transition, affects the crystalline lattice or zero-point fluctuations.

Although the lattice parameters for a given crystallite remained fairly stable as a function of temperature, due to different annealing conditions and difficulties with temperature stability between different runs, there were some changes in the values obtained. These changes had little impact on the observed mean-square displacement, indicating that we had observed the temperature- and volume-independent zero-point motion here.

## 5. Conclusions

In summary, we have obtained microscopic quantitative data such as mean kinetic energy, mean square momentum and mean square displacement of helium atoms, as well as the lattice parameter for solid $^4$He in temperature range 70mK – 500 mK. We did not see any change in the single atom kinetic energy (within 1%) or change in the lattice parameter (within 0.03%). The mean square displacement did not change in the region of expected supersolid transition either. All these results suggest that NCRI transition could be different to the superfluid transition in liquid $^4$He.

## 6. Acknowledgment

The author would like to thank J. M. Goodkind, H. R. Glyde, M. A. Adams, J. Mayers, C. Broholm, E. Blackburn, S. K. Sinha, J. Hudis, S. O. Diallo, J.V. Pearce, R. T. Azuah, J. van Duijn, C. D. Frost, J. W. Taylor, and R. B. E. Down for the fruitful collaboration and the staff at the ISIS Facility for their assistance with the low temperature and high pressure equipment needed.